\documentstyle[amssymb,aps,prb,preprint]{revtex}

\begin{document}
\title{Low temperature specific heat of the molecular cluster Fe$_{8}$:
contribution of the local field}
\author{A. M. Gomes, M. A. Novak, W. C. Nunes and R. E. Rapp}
\address{Inst. de F\'{i}sica - Universidade Federal do Rio de Janeiro\\
Caixa Postal 68528\\
Rio de Janeiro, \\
21945-970, Brazil}
\maketitle

\begin{abstract}
We present low temperature specific heat and ac-susceptibility measurements
on Fe$_{8}$ powdered sample. Below 1.3 K, superparamagnetic blocking effects
as well as an excess specific heat contribution are evident. The latter is
attributed to a splitting of the ground state doublet in an inhomogeneous
local field of hyperfine and dipolar origin. The local field contributions
are evident in the resonances observed in the field dependent
ac-susceptibility and specific heat below 0.5 K. The low temperature
Schottky contribution allowed us to check the crystal field parameters of
effective spin Hamiltonians, recently proposed to simulate EPR and inelastic
neutron scattering experiments.
\end{abstract}

\section{Introduction\protect\smallskip}

Recently nanomagnetic molecular cluster compounds have attracted attention
due to the clear demonstration of collective quantum effects and its
implications in information storage and quantum devices. Compounds based on
organometallic clusters have been studied in physics and chemistry as its
magnetic properties can be modified by manipulation of its molecular
structure.\cite{Gatteschi} Some of these clusters present a high spin ground
state with high anisotropy. At low temperatures, it was found that
magnetization reversal occurs by quantum tunneling through the anisotropy
energy barrier, becoming a prototype system to study macroscopic quantum
tunneling (MQT). \cite{Leggett,Barbara} This was possible as these clusters
form a crystal of identical and non interacting macrospins with magnetic
properties easily detected. Although this system cannot be directly related
to a macroscopic entity, as its typical size is of a few angstroms, their
magnetic properties are governed by the magnetic moment associated to the
total spin of the cluster, which is in the borderline of the mesoscopic
scale. The most studied molecular cluster of this kind is Mn$_{12}$Ac, which
has shown spectacular effects such as steps in magnetic hysteresis due to
thermally activated field resonant tunneling.\cite{FriedmanThomas} The
crossover temperature to pure quantum regime, where the magnetic relaxation
becomes temperature independent is not well established yet.\cite
{PaulsenPerenboom} On the other hand, the theory developed explains
reasonably most effects found for this material\cite
{Politi,Fort,Zvezdin,Garanin,Villain,Prokofiev,Leuenberger} although some
features are not yet explained,\cite{NovakAngelo} specially the second
relaxation process and the excess specific heat found at low temperatures.

The Fe$_{8}$ molecule was the first nanomagnet to show clear evidence of a
pure quantum tunneling regime at low temperatures around 0.4 K.\cite
{Sangregorio} Having a smaller energy barrier, 28-32 K, as compared to $%
\thicksim $70 K in Mn$_{12}$Ac, and the considerable transversal anisotropy,
it became the most appropriate system to study quantum tunneling of
magnetization (QTM). Very recently, this system allowed the experimental
observation of spin quantum phase interference effects.\cite{Wernsdorfer}

Here, we present absolute measurements of specific heat and
ac-susceptibility in this octonuclear iron cluster. The novelty in our
studies of this compound is the role played by the interaction of the spins
with the internal local fields on the specific heat results at very low
temperatures. Our data show that there is a considerable contribution that
can be well explained in terms of the already known Hamiltonian and an
excess contribution due to a distribution of local internal fields.

\section{The Fe$_{8}$ cluster}

The Fe$_{8}$ molecular cluster was reported to be synthesized for the first
time in 1984, its chemical formula is [Fe$_{8}$O$_{2}$(OH)$_{12}$(tacn)$_{6}$%
]$^{8+}$, where tacn is the tryazacyclononane molecule.\cite{Wieghardt} The
iron atoms are embedded in a organic matrix, where four iron III are bridged
by hydroxo groups to central iron atoms. A preliminary study has reported an
uncompensated antiferromagnetic ground state with spin S=10, presenting a
superparamagnetic behavior.\cite{Delfs} The cluster has approximately a D$%
_{2}$ symmetry and a proposed structure,\cite{Delfs,Laura} of six irons III
ions (S=5/2) pointing in opposition to the remaining two, recently confirmed
by polarized neutron experiments.\cite{Pontillon} At low temperatures, the
magnetic relaxation of this cluster follows roughly an Arrhenius law with
energy barrier of approximately 28 K,\cite{Laura} in fair agreement with
high field EPR spectra experiments showing an easy-axis with strong in-plane
(rhombic) anisotropy.\cite{Laura} The effective spin Hamiltonian (EPR) used
to simulate the results is given by:

\begin{equation}
\varkappa =D[S_{z}^{2}-S(S+1)/3]+E(S_{x}^{2}-S_{y}^{2})+H^{\prime }\text{ ,}
\label{Hamold}
\end{equation}
where $D=-0.27$ K , $E=0.046$ K and H' is the Zeeman splitting term. The
transversal anisotropy term in this cluster is different and considerably
larger than in Mn$_{12}$Ac being responsible for the clear observation of
quantum tunneling of magnetization in this system.

Recently, inelastic neutron spectroscopy of Fe$_{8}$ single crystals has
shown that a more complex Hamiltonian\cite{Caciuffo} (INS) is necessary to
describe the cluster energy levels:

\begin{eqnarray}
\varkappa &=&D[S_{z}^{2}-S(S+1)/3]+E(S_{x}^{2}-S_{y}^{2})+D^{\prime }\hat{O}%
_{4}^{0}(S)+  \label{Hamnew} \\
&&+E^{\prime }\hat{O}_{4}^{2}(S)+C^{\prime }\hat{O}_{4}^{4}(S)+H^{\prime }%
\text{ ,}  \nonumber
\end{eqnarray}
where the expressions for $\hat{O}_{4}^{0}(S)$, $\hat{O}_{4}^{2}(S)$, $\hat{O%
}_{4}^{4}(S)$ contain operator in $S_{z}^{n}$, $S_{+}^{n}$ and $S_{_{-}}^{n}$
($n=2,4$). The complete expressions can be found in Ref. 21 with $D=-0.29$
K, $E=0.047$ K, $D^{\prime }=1.00\times 10^{-6}$K, $E^{\prime }=1.16\times
10^{-7}$ K , and $C^{\prime }=8.56\times 10^{-6}$ K, corresponding to a
total energy barrier of 32.8 K. Note that S$_{z}$ does not commute with $%
\varkappa $, the pairs of $+m_{s}$ states are splitted and the energy levels
are related to a mixture of different $m_{s}$ states.

Magnetic relaxation curves for this cluster show for temperature above 1.3 K
a single exponential relaxation, while for lower temperatures the system
presents a more complex behavior described by a stretched exponential.\cite
{Sangregorio} Temperature independent relaxation was observed below 0.4 K
indicating that a clear quantum tunneling regime is attained below this
temperature. Above 1.5 K ac-susceptibility and M\"{o}ssbauer spectroscopy
experiments\cite{Laura} show that this system follows a thermal activation
law:

\begin{equation}
\tau =\tau _{0}e^{\frac{\Delta E}{k_{b}T}}\text{,}  \label{Arrhenius}
\end{equation}
with $\tau _{0}=3.4\times 10^{-8}$ s and $\Delta E/k_{B}=24.5$ K, an energy
barrier about three times smaller than in Mn$_{12}$Ac. This fact in
conjuction with the strong rhombic term allowed the clear observation of
magnetic relaxation in the quantum regime in the low temperature limit.

\section{Specific heat measurements}

Calorimetric measurements are useful to check the energy levels obtained
from Hamiltonians (Eq. \ref{Hamold} and \ref{Hamnew}), as was recently done
for Mn$_{12}$Ac.\cite{NovakAngelo} For Fe$_{8}$, in contrast with Mn$_{12}$%
Ac, hyperfine contribution should be almost negligible, as natural abundance
of $^{57}$Fe is 2.2\%, allowing an easier separation of the different
contributions to the specific heat at very low temperature.

Specific heat measurements between 1.3 and 20 K were made with an automated
semi-adiabatic calorimeter in a pumped He$^{4}$ refrigerator while
measurements below 1.5 K were done in a similar calorimeter within a
dilution refrigerator. We measured 293 mg of Fe$_{8}$ powder sample mixed
with copper powder and Apiezon N grease in order to increase thermal
contact. Data is obtained by monitoring the temperature time dependence of
the sample assembly before and after a heat pulse of about 4 minutes is
applied to the calorimeter. Correction for the addenda was previously
obtained representing about 40 \% of the total heat capacity in the worst
case. The whole specific heat curve from 80 mK up to 22 K is shown in the
inset of Fig. 1, with an overall experimental error being less than 4\%.
Results do not present the unexplained small anomalies reported recently by
Fominaya et al.\cite{Fominaya} around 2 K. Their results were obtained by a
different technique, the ac method, using a unique single crystal. Our data
obtained by the two measurement systems are superposed between 1.3 and 1.5 K
with a discrepancy smaller than 1\%. Data below 2 K (see Fig. 1) reveals two
surprising features: the specific heat curve presents an expressive
reduction around 1.2 K and an upturn as temperature decreases. At first,
this Schottky-like contribution could be thought to be originated from
hyperfine and/or superhyperfine interactions with other nuclei in the
cluster, but this should be relevant only below 200 mK. A more realistic
possibility is due to Zeeman splitting of the ground state by internal
dipolar fields. Both hypothesis will be discussed latter in the text.

We analyzed the experimental data calculating the zero field splitting
Schottky (ZFS-Schottky) using the two proposed Hamiltonians (EPR and INS), .
For zero internal magnetic field $H^{\prime }=0$, only the crystal field
energy levels are calculated by diagonalization of the ($2S+1$) $\times $ ($%
2S+1$) matrix using only $S=10$ multiplet and the $S_{z}$ eigenstates.
Specific heat contribution is calculated numerically using the eigenvalues $%
\varepsilon _{i}$ of the diagonalized matrix and differentiating the
expression for the mean energy of the system:

\begin{equation}
c=\frac{d\overline{E}}{dT}=\frac{d}{dT}\left( R\frac{\sum%
\limits_{i=-m_{s}}^{m_{s}}\varepsilon _{i}\exp (-\varepsilon _{i}/T)}{%
\sum\limits_{i=-m_{s}}^{m_{s}}\exp (-\varepsilon _{i}/T)}\right) \text{,}
\label{Emed}
\end{equation}
where $R=8.314$ J$\cdot $K$^{-1}$. The calculated curves using both
Hamiltonians are shown in Fig. 2. As can be seen, the INS and EPR
Hamiltonians produce slightly different results, indicating that the
forth-order terms have a small but not negligible influence in the specific
heat. Data below 1 K cannot be explained by both Hamiltonians only, the
calculated contributions decreases monotonically, becoming zero below 0.5 K,
whereas the experimental data show an increase.

Hyperfine contribution due to $^{57}$Fe isotopes should be negligible. This
contribution yields a T$^{-2}$ term that is observable only at very low
temperatures. An estimate of the hyperfine term was made for Fe$_{10}$ and Fe%
$_{6}$ samples by Affronte et al.\cite{Affronte,Affronte2} and could be
applied to Fe$_{8}$. In this case, the coefficient of the T$^{-2}$ term is
of the order of $10^{-6}$ J K mol$^{-1}$ and cannot explain the observed
increase of the specific heat at low temperatures.

Measurements with an applied dc-magnetic field were made within a
superconducting solenoid in the temperature range between 1.3 and 23 K. The
results are shown in Fig. 3 for 0.17 T and 0.32 T being similar though more
pronounced than those obtained for Mn$_{12}$Ac.\cite{NovakAngelo,Gomes}
Here, a gradual decrease of the specific heat can be observed for
temperatures below 1.6 K. Below 1 K the relaxation times becomes bigger than
1 minute and cease to contribute to the specific heat measurements, being in
the blocked regime. The field effect is better seen in the inset, where the
data is plotted as C/T versus T. The calculation with both Hamiltonians were
repeated, including now a Zeeman term $H^{\prime }=g\mu SH$, and performing
a powder average in the mean energy expression, with a discrete sum over $%
\theta $, the angle between the magnetic field and the easy axis:

\begin{equation}
\overline{E}=\left( R\sum\limits_{\theta =0}^{\pi }\frac{\sum%
\limits_{i=-m_{s}}^{m_{s}}\varepsilon _{i}\exp (-\varepsilon _{i}/T)}{%
\sum\limits_{i=-m_{s}}^{m_{s}}\exp (-\varepsilon _{i}/T)}\sin \theta \right)
/\sum\limits_{\theta =0}^{\pi }\sin \theta \text{ ,}  \label{Emedpo}
\end{equation}

In Fig. 3, the calculated Schottky contribution for 0.17 T is plotted using
Eq. 5. In this figure, the lowest temperature of the equipment unfortunately
limits the experimental data under external magnetic field, not available in
the dilution refrigerator system. Even so, a downturn can be clearly seen
below 2 K. This can be explained again as due to superparamagnetic blocking
in the time scale of the specific heat measurements. At higher temperatures,
above 10 K, the Zeeman contribution, for those small fields, becomes
negligible, not affecting the specific heat results.

\section{Ac-susceptibility and local field distribution}

We have performed ac-susceptibility measurements with a Fe$_{8}$ powder
sample from 10 Hz to 10 kHz and temperatures from 2.4 to 15 K. In Fig. 4 is
shown the plot of the real and imaginary components of the ac-susceptibility
as function of a static external magnetic field.

The oscillating field was of the order of 0.1 mT at 10 kHz. Above the
blocking temperature (for 10 kHz, T$_{B}$=3.5K) several dips are observed in
the imaginary component, while the real one is smooth. At the measured
temperatures, $T=6.6$ and $7.1$ K, a large part of the relaxation mechanism
is dominated by thermal activation and thermally activated tunneling process
near the top of the barrier. The dips occur at the positions where steps
were found in the magnetization hysteresis cycles.\cite{Sangregorio} Below T$%
_{B}$, the real component of the susceptibility also show maxima. These are
attributed to level crossing fields and are not regularly spaced in field as
indicated in the Fig. 4. The important point here is the width of the dips,
which is the same as the width of the maxima observed in the real component
at lower temperatures. The expected width of the tunneling resonance (tunnel
splitting) was estimated by Prokofe'v and Stamp to be only $10^{-8}$ mT for
Mn$_{12}$Ac, for Fe$_{8}$ it is expect also to be very small, not explaining
the observed resonance width. This is a clear indication that there is a
distribution of \ local fields mainly of dipolar origin superposed with
hyperfine fields from the few $^{57}$Fe, N, and H nuclei present in the
molecule. The dipolar local field distribution was also probed recently by
Wernsdorfer et al.\cite{Wernsdorfer2} which found a width ($\sigma _{d}$) of
the order of 50 mT, in agreement with estimates for dipolar contribution of
the surrounding neighbors. Data in Fig. 4 have a width of 0.11 T, which is
larger than the ones reported in Ref. 23, due to the fact that a powder
sample was used.

These internal fields have been already addressed\cite{Ohm,Wernsdorfer3} in
order to explain the magnetic relaxation process. The spins which are within
the narrow resonant condition, at some specific local field, tunnel
producing spin flips which in turn change the neighboring dipolar field,
bringing up other clusters into resonance resulting in a complex evolution
of the internal fields and relaxation. We discuss now the influence of this
local field on the specific heat results. For simplicity, we assume that
each cluster spin is in a local field with a probability given by the field
distribution $P(H)$ independent of the temperature. The specific heat ($%
c^{\prime }$) is obtained by averaging over the distribution $P(H)$: 
\begin{equation}
c^{\prime }(T)=\frac{%
\mathop{\displaystyle \sum }%
\limits_{i}c_{i}(T_{,}H_{i})P_{i}(H_{i})}{%
\mathop{\displaystyle \sum }%
\limits_{i}P_{i}(H_{i})}\text{ ,}  \label{cprob}
\end{equation}
where $c_{i}(T,H_{i})$ is the calculated specific heat at a field $H_{i}$.
We have used both Lorentzian and Gaussian distributions centered at $H=0$
and width 0.1$<\sigma _{d}<$ 70 mT. The calculated specific heat for both
distributions present no quantitative difference for temperatures above 2 K.
However there is a slight difference between them below 2 K. In Fig. 5 we
show in more detail the results below 2 K and the different contributions
important in this range. It is clear that by using a Lorentzian distribution
with $\sigma _{d}=$10 mT, the calculated curve lies well above the
experimental results below 1 K, though presenting the same shape. The other
important feature is the decrease in the experimental data observed below
1.3 K, which is the superparamagnetic blocking temperature observed in
magnetic relaxation experiments. Similar effect was found previously in
specific heat measurements on Mn$_{12}$Ac powder sample \cite{Gomes} under
small external magnetic field. This is another evidence that in Fe$_{8}$
blocking effects are observed due to a splitting of the ground state by
internal fields. Well below this blocking temperature, this contribution
should vanish unless part of the clusters may tunnel to the other side of
the anisotropy barrier.

Using a field distribution larger than 10 mT produces a specific heat
maximum above 0.1 K, being contrary to experimental results. As  $\sigma _{d}
$ increases this maximum shifts to higher temperatures, acquiring a
different shape from the experimental data. Thorsten et al.\cite{Ohm} had
succeeded to describe the fast relaxation process in Fe$_{8}$ with a
Gaussian distribution of dipolar fields with $\sigma _{d}=$ $40$ mT. On the
other hand, using a phenomenological model that accounts for the evolving
distribution of local dipolar fields in time they found $\sigma _{d}=$ $10$
mT distribution. Both cases do not explain fully our data, and indicate that
the hyperfine fields have to be taken into account.

We calculated this hyperfine contribution \ by assuming that each Spin 10 is
subjected to a field following a Lorentzian distribution of width $\sigma
_{h}=$ 2 mT, shown separately in Fig. 5. This is in fair agreement with
estimates made by Wernsdorfer et al.\cite{Wernsdorfer3} which found the line
width of the hyperfine field fluctuations to be about 1.2 mT. Even being
much higher than pure $^{57}$Fe hyperfine contribution discussed previously,
it is still too small to explain the experimental results. In order to
explain the full curve one has to take into account all those contributions
and the blocking effect, which reduce them. This is accomplished if we
assume that a fraction of the total clusters are unblocked during the course
of the experiment. Note that below 0.5 K the magnetic relaxation is in the
quantum regime, being temperature independent. The blocked clusters do not
contribute to the specific heat in this temperature range, only the clusters
that change its magnetic orientation either by thermal activation and/or
tunneling, in a time shorter than the measuring time. The data was only well
fitted when considering 30\% of both dipolar and hyperfine contributions as
shown as a full line in Fig. 5.

Finally we now discuss the phonon contribution to the specific heat. This
can be found by analyzing the data above 2 K after subtracting the magnetic
contributions, as displayed in Fig. 6 in the traditional $c\cdot
T^{-1}\times T^{2}$ plot. Here we also took into account the tail of the
local field contribution, which depends on its width. We have found that
there is an apparent linear term in the specific heat which depends
considerably on the local field distribution width. This was already
observed on Mn$_{12}$Ac with no explanation.\cite{NovakAngelo,Gomes} By
using the 10 mT width distribution, the linear term is practically nulled as
showed in Fig. 6. From the slope of the fitted straight line the Debye
temperature $\ \theta _{D}=$($33\pm 1$) K is determined.

\section{CONCLUSIONS}

Specific heat of the nanomagnet Fe$_{8}$ was measured down to 0.1 K. For
temperatures above 1 K, the results are dominated by the crystal field
Schottky anomaly and lattice contributions and the recently proposed
Hamiltonian used to fit inelastic neutron scattering experiments gives the
best agreement with our data. Below 1 K an excess contribution associated to
a distribution of local fields of dipolar and hyperfine origin is evidenced.
The results indicate that dynamical hyperfine and dipolar fields play a
significant role to explain the specific heat below 0.6 K and a good
agreement between the calculated contributions and the experimental data is
obtained only when we consider a fraction of the spins remain unblocked, due
to magnetic tunneling, below the known superparamagnetic blocking
temperature.\newline

\begin{center}
{\bf ACKNOWLEDGMENTS}
\end{center}

We would like to acknowledge D. Gatteschi and coworkers for supplying us
with the sample and R. Sessoli, C. Sangregorio and W. Wernsdorfer. for
useful discussions. We would like to acknowledge also CNPq, FUJB for
financial support.

FIGURE\ CAPTIONS\newline

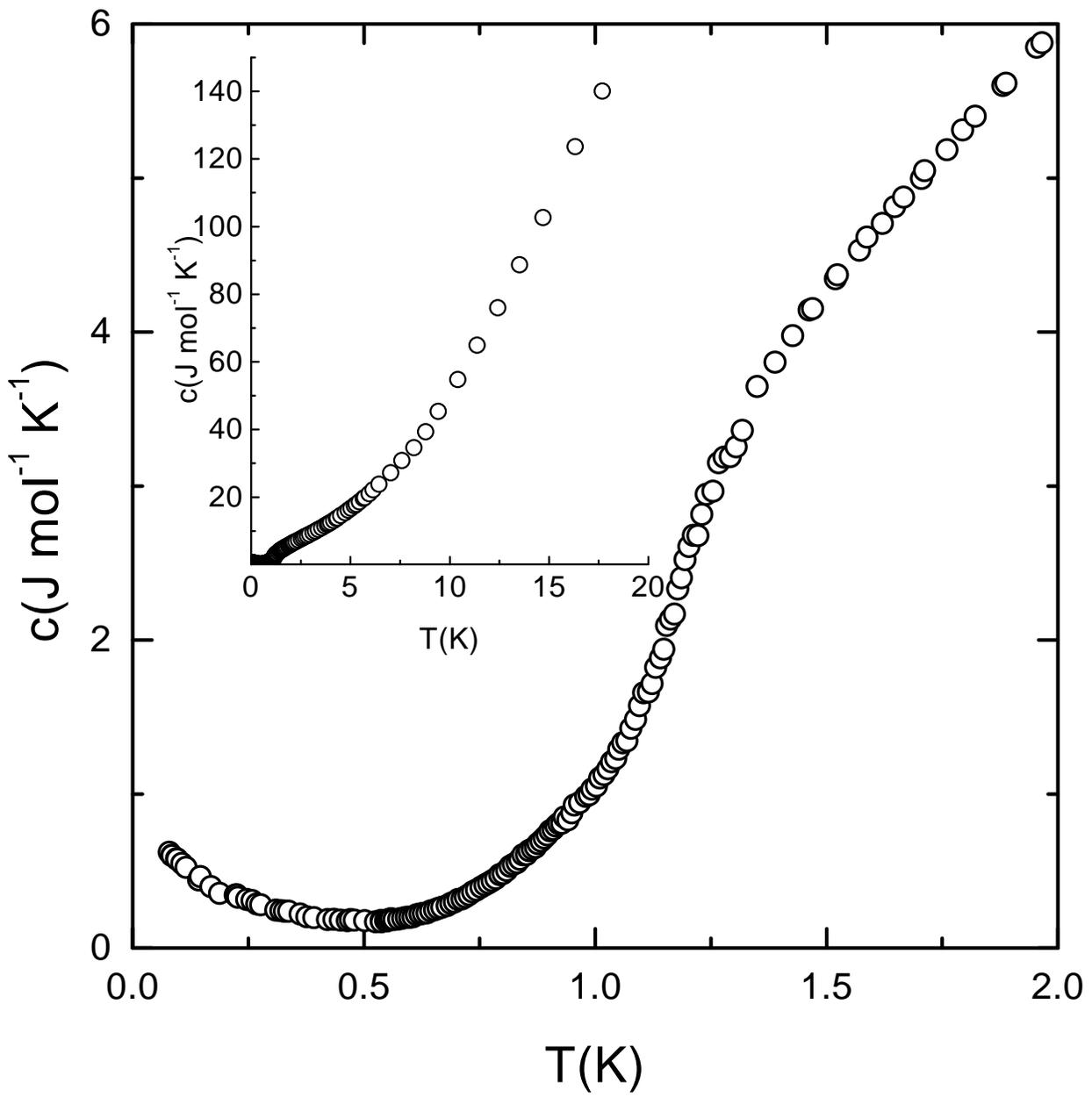
FIG. 1. Experimental specific heat results for Fe$_{8}$ between $80$ mK $%
<T<2 $ K. Data decreases abruptly below 1.5 K and at lower temperatures an
excess specific heat is present. Inset: complete specific heat curve up to
18 K.\newline

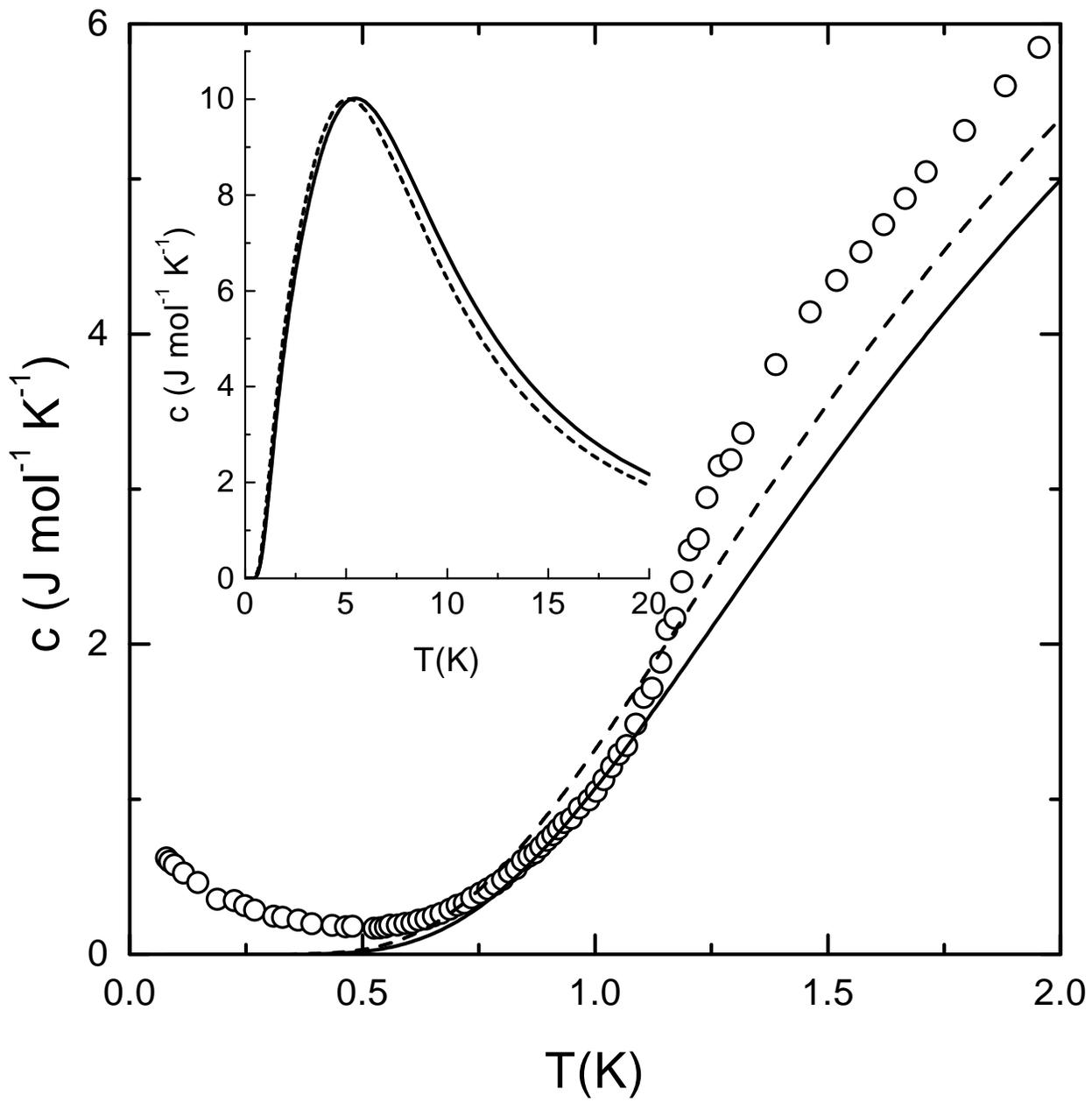
FIG. 2. Calculated ZFS-Schottky for EPR and INS Hamiltonians. Open circles:
experimental data. Dotted line: EPR Hamiltonian. Full line: INS Hamiltonian.
Inset: calculated curves up to 20 K.

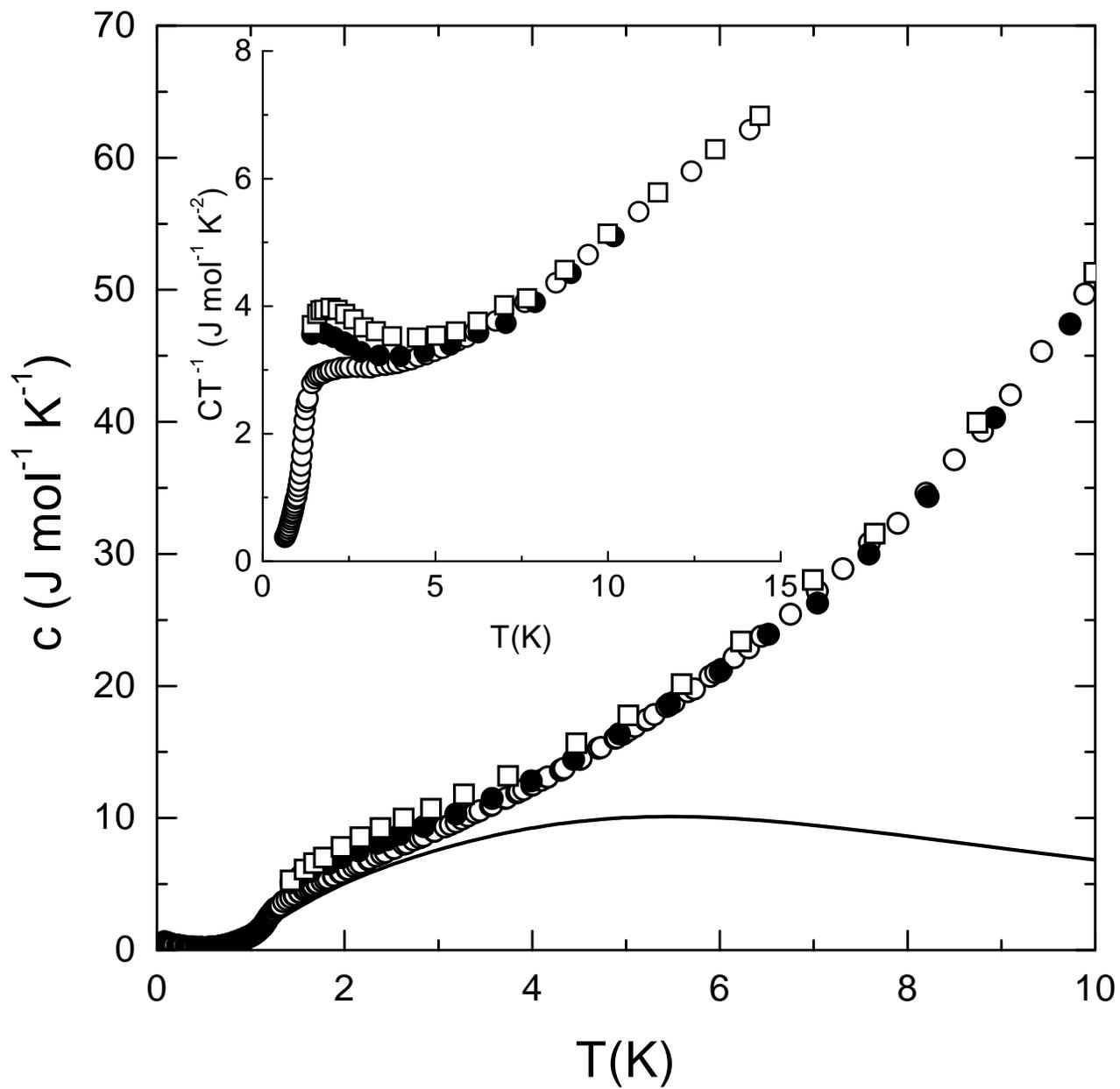
FIG. 3. Low temperature specific heat in external applied magnetic field of
0.17 T (full circles), and 0.32 T (squares) compared to the experimental
data with $H^{\prime }=0$ (open circles). Inset: Same data shown in a $%
c\cdot T^{-1}\times T$ plot where the onset of the superparamagnetic
blocking effect in the specific heat can be seen.\newline

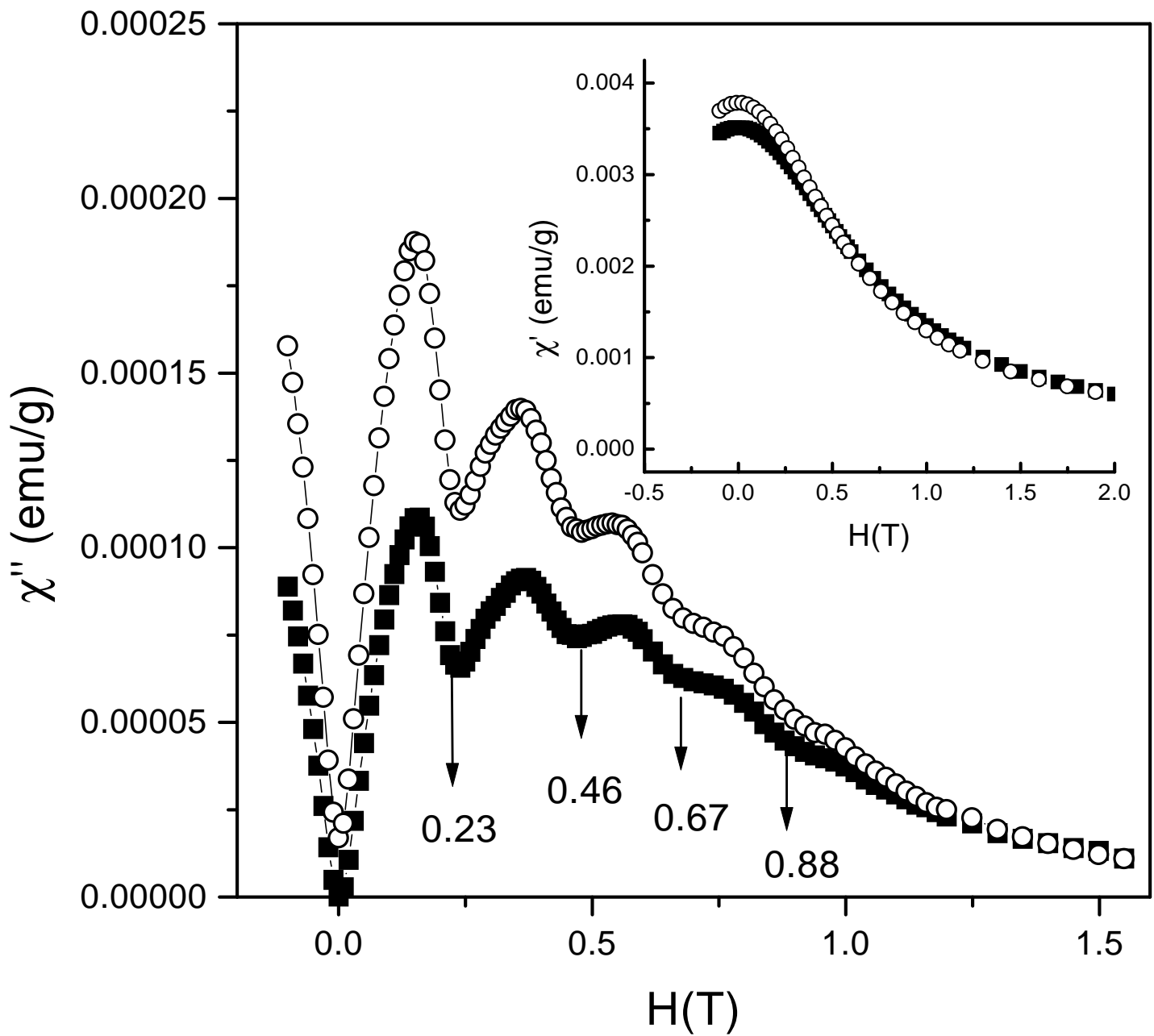
FIG. 4. Imaginary component of the ac-susceptibility as function of the
external magnetic field for 10 kHz at 7.1 and 6.6 K, full squares and open
circles respectively. Resonant fields are identified by a minima in the
curves ($H=$ 0, 0.23, 0.46, 0.67 and 0.88 T). Inset: real component which
does not show any resonance.\newline

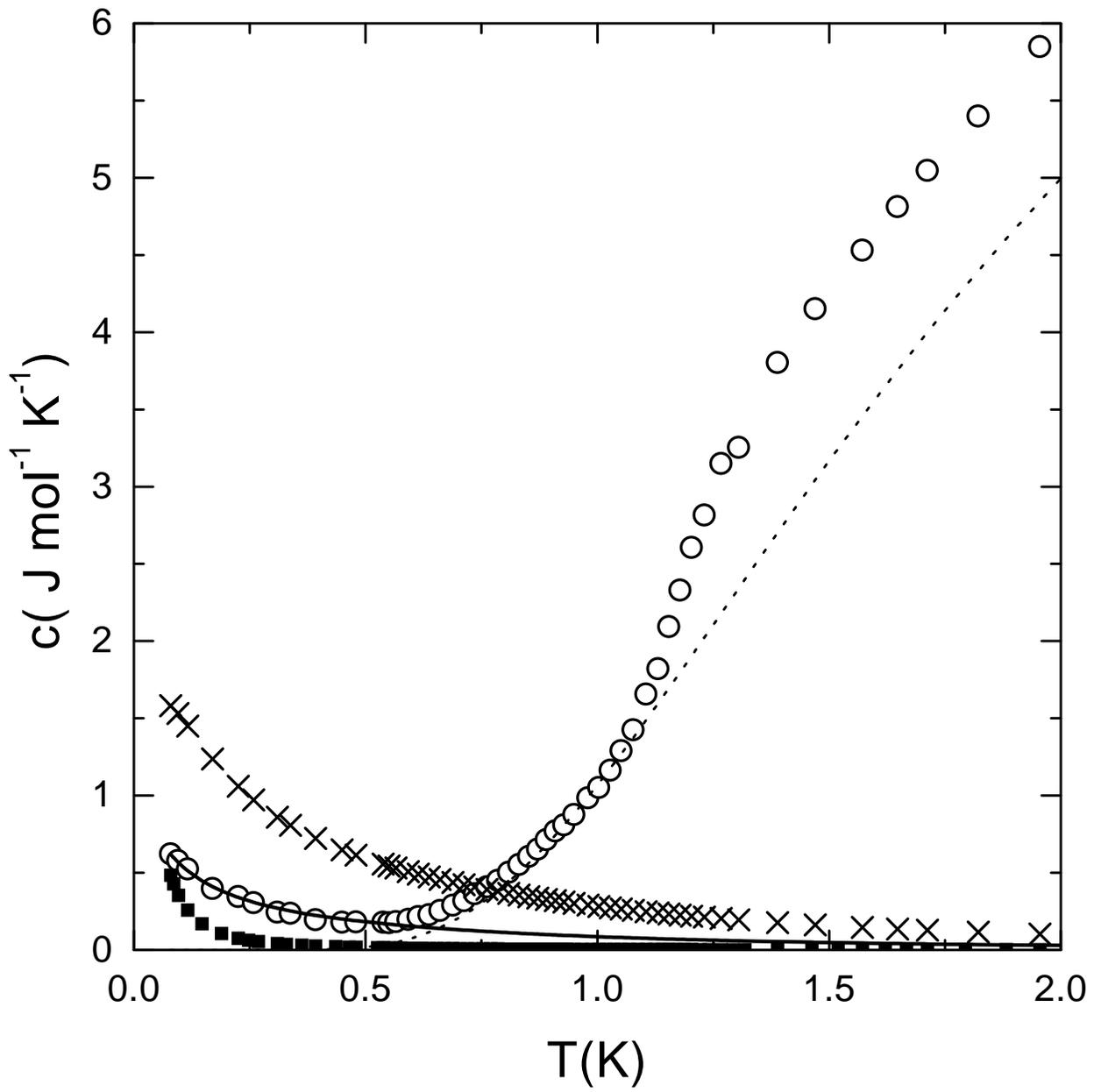
FIG. 5. Experimental data (empty circles) below 0.5 K is fitted by assuming
30\% (full line) local dipolar contribution with $\sigma _{d}=10$ mT
(crosses) and hyperfine distribution with $\sigma _{h}=2$ mT (full squares).%
\newline

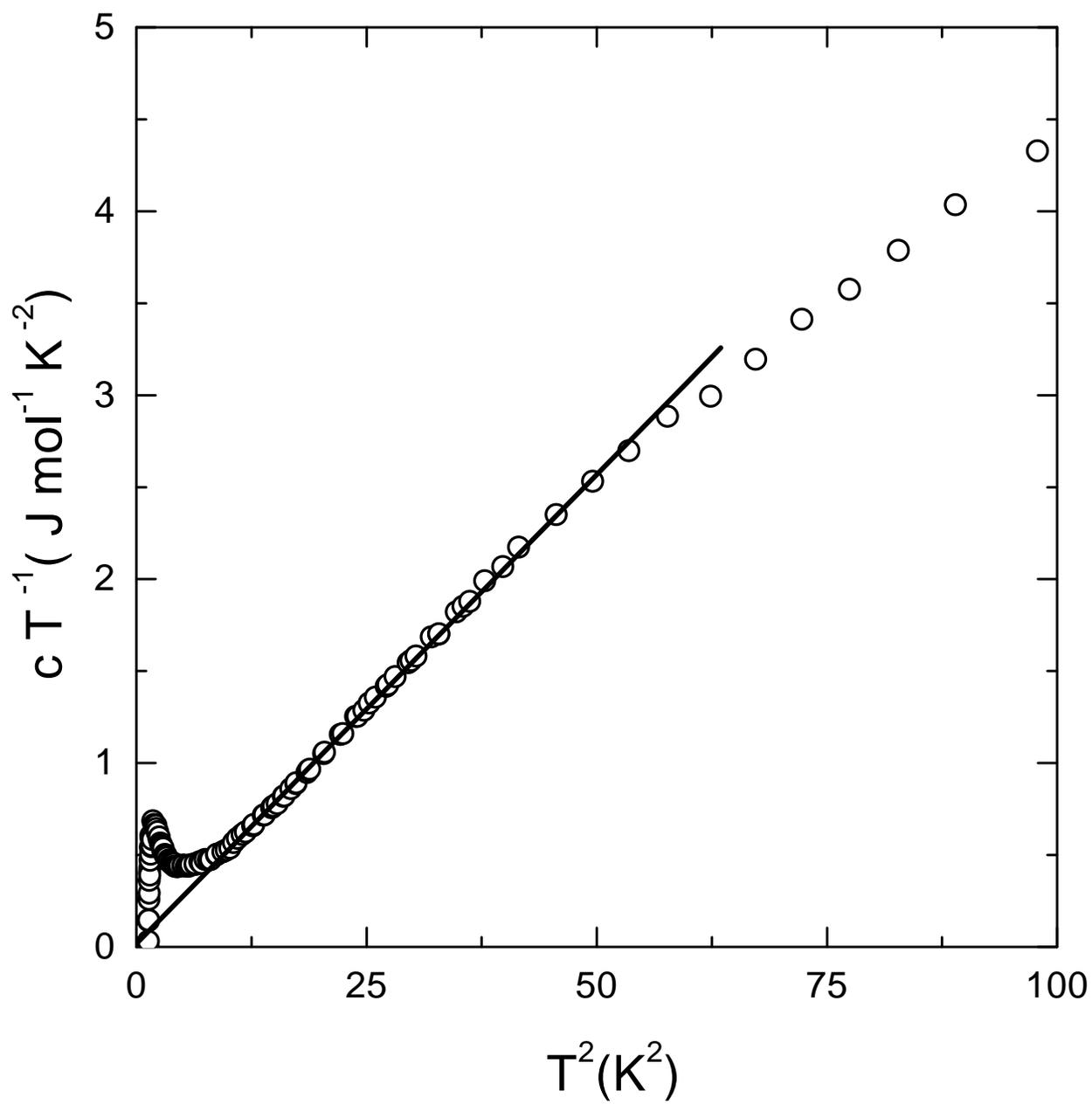
FIG. 6. $c\cdot T^{-1}\times T^{2}$ plot after subtracting the local dipolar
field contribution. The Debye temperature, $\theta _{D}=(33\pm 1)$ K, is
obtained from the straight line fit for data above the blocking temperature.%
\newline


\begin{references}
\bibitem{Affronte2}  Note that in Ref. 25 the correct value for the T$^{-2}$
term of the hyperfine contribution is 10$^{-2}$ times smaller than the
printed in the text, even so, it does not change the authors conclusions.
\end{references}
\end{document}